\documentclass[12pt,preprint]{aastex}

\shorttitle{The Sun's Open Magnetic Field}
\shortauthors{Antiochos et al.}

\begin{document}

\title{Structure and Dynamics of the Sun's Open Magnetic Field}

%% You can use \email to mark an email address anywhere in the paper.

\author{S. K. Antiochos, C. R. DeVore, and J. T. Karpen}
\affil{Naval Research Laboratory, Washington, DC, 20375}
\email{antiochos@nrl.navy.mil}

\and

\author{Z. Mikic}
\affil{Science Applications International Corporation, San Diego, CA 92121}

%% title and affiliation information. 

\begin{abstract}

The solar magnetic field is the primary agent that drives solar
activity and couples the Sun to the Heliosphere. Although the details
of this coupling depend on the quantitative properties of the field,
many important aspects of the corona - solar wind connection can be
understood by considering only the general topological properties of
those regions on the Sun where the field extends from the photosphere
out to interplanetary space, the so-called open field regions that are
usually observed as coronal holes. From the simple assumptions that
underlie the standard quasi-steady corona-wind theoretical models, and
that are likely to hold for the Sun, as well, we derive two
conjectures on the possible structure and dynamics of coronal holes:
(1) Coronal holes are unique in that every unipolar region on the
photosphere can contain at most one coronal hole. (2) Coronal holes of
nested polarity regions must themselves be nested.  Magnetic
reconnection plays the central role in enforcing these constraints on
the field topology. From these conjectures we derive additional
properties for the topology of open field regions, and propose several
observational predictions for both the slowly varying and transient
corona/solar wind.

\end{abstract}



%% notice the use of the natbib \citep

\section{Introduction}

Decades of solar XUV/X-ray observations have shown that the Sun's
corona is divided into two distinct types of magnetic regions: those
in which field lines are ``closed'', i.e., rooted to the photosphere
at two ends, and those in which field lines are ``open'', rooted to
the photosphere at only one end and extending out to the
Heliosphere. The fast solar wind is believed to emanate from these
open field regions, which usually appear dark in X-rays and,
therefore, are referred to as coronal holes.  Although a great deal of
progress has been made on understanding the Sun's open magnetic field
since the seminal work of \citet{parker58}, many questions remain
outstanding, especially concerning the dynamics of solar-heliospheric
coupling. For example, we still do not understand the magnetic
topology and evolution that give rise to the slow wind
\citep[e.g.,][]{axford97, zurbuchen07}. At an even more basic level,
the coronal and heliospheric observations themselves appear to be in
conflict. Coronal images often show streamer evolution and
inflows/outflows that are interpreted as closed flux opening into the
wind and open flux closing back down to the streamer
\citep{hundhausen84, howard85, sheeley02}. Furthermore, the
long-standing observation that some coronal holes appear to rotate
rigidly, impervious to differential rotation \citep{timothy75,
zirker77}, implies the continuous opening and closing of coronal field
\citep{wang96, lionello06}. However, the opening of closed flux
implies the injection of field lines into the wind with both
footpoints connected to the photosphere. Such field lines should
exhibit a bi-directional heat flux (counter-streaming electrons) in
heliospheric measurements, yet counter-streaming electrons are rarely
seen outside of interplanetary coronal mass ejections (ICMEs)
\citep{gosling90}. Conversely, the closing down of open flux would
imply the formation of U-shaped field lines in the heliosphere with no
connection to the Sun, resulting in a heat flux drop-out in the
heliospheric measurements \citep{mccomas89, mccomas91, lin92}. Again,
these are rarely seen outside of ICMEs \citep{pagel05}. Apparently,
the heliosphere does not care what the corona is doing!

This apparent contradiction among the observations has led to
conflicting approaches to modeling the topology and evolution of the
open field. Motivated by their observations, some heliospheric
researchers have proposed the interchange model, in which reconnection
between open and closed flux is the dominant process by which the
coronal open field evolves \citep{fisk99, fisk05, fisk06}. Note that
reconnection between an open and a closed field line produces another
pair of open and closed field lines, and no disconnected lines
\citep{crooker02}. In this model the magnetic topology is presumed to
be highly complex, with open field mixing into closed field regions as
in a diffusion process.

On the other hand, motivated by their observations solar researchers
have primarily used the quasi-steady model in which the coronal
magnetic field is calculated from the instantaneous normal component
at the photosphere using some extrapolation procedure.  The simplest
and most widely-used such procedure is the source surface model, in
which the field is assumed to be current-free in the corona and to be
purely radial at some fixed spherical surface \citep{altschuler69,
schatten69, hoeksema91}. This model predicts a topology consisting of
distinct open flux (coronal holes) and closed flux regions
\citep{wang90, arge00, luhmann02, luhmann03, schrijver03}. Although
the dynamics are not explicitly calculated, field-line opening and
closing driven by the evolution of the photospheric flux are implicit
in this model. The assumption is that the coronal field evolution can
be approximated by a sequence of source-surface solutions. The
quasi-steady models have been extended in recent years to include
solution of the full MHD equations, no longer requiring the assumption
of a current-free magnetic field in the corona and an artificial
source surface. Typically, the MHD models are used to compute a
steady-state equilibrium \citep[e.g.,][]{linker99, riley01,
odstrcil03, roussev03}, but they can also be employed for fully
dynamic simulations \citep{riley03}.

It should be emphasized that the quasi-steady models are nothing more
than particular implementations of Parker's basic theory of the solar
wind.  In his seminal work, Parker (1958) argued that if the gas
pressure in the corona becomes larger than the magnetic pressure, the
gas must expand outward as a wind, dragging the field lines with
it. Hence, the open and closed regions are fundamentally determined by
the magnitude of coronal heating and the magnitude of the field in the
corona. Implicit in Parker's theory is the assumption that field will
open and close in response to changes in these quantities. In
principle, the MHD models are a more physical representation of the
theory; but since the coronal heating mechanism is not known, they are
also {\it ad hoc} in practice.

Although the interchange and quasi-steady models predict very
different topology and dynamics of the open field, they both agree
that reconnection is the dominant physical process involved in the
opening and closing of magnetic flux. Reconnection is also the process
generally invoked for transient opening and closing of flux by CMEs
\citep[see reviews by][]{forbes00, klimchuk01, low01, wu01, linker03,
lin03}.  In this paper we adopt the solar-centric view that the
coronal field is given by some straightforward extrapolation of the
photospheric boundary conditions, such as a source-surface or MHD
solution.  Using two straightforward assumptions and some analytic
arguments, we derive strict constraints on the open-field topologies
allowed by the quasi-steady models. We also discuss the central role
of reconnection in determining coronal hole structure and dynamics.

\section{Basic Assumptions of the Quasi-Steady Model}

The first assumption is that all rapidly varying, small-scale
structure in the photospheric field, such as the clumping into
elementary flux tubes \citep{solanki93, muller94} and the so-called
magnetic carpet \citep{schrijver97}, can be neglected when calculating
the topology of the open field. It should be emphasized that, given
sufficient numerical resolution, the quasi-steady models can
incorporate arbitrary spatial structure in the photospheric flux
distribution. The influence of the small-scale photospheric structure
generally disappears at heights far below the source surface; but, if
desired, one could always include this spatial structure in the
model. The problem, however, is that the small-scale structure may not
be compatible with the assumption of quasi-steady evolution.  Since
both the magnetic field and the wind must reach a steady state, the
quasi-steady assumption requires an evolution slow compared to either
a sound or Alfven crossing time of the corona. Hence for a typical
sound speed of $\sim 10^7$ cm/s and scale of order the solar radius
$\sim 10^{11}$ cm, the restriction is that the photospheric field
evolves on the time scale of hours or longer.

This restriction holds for fields of the scale of active regions, but
not for the small magnetic-carpet bipoles. The effect of these small
temporal and spatial scale bipoles is to add some high-frequency
``noise'' to the system. The interchange advocates would argue that it
is exactly this ``noise'' that is responsible for the distinguishing
feature of their model: open field can diffuse freely into closed
field regions.  It seems unlikely, however, that the small-scale
carpet fields can have such a dramatic effect on the field
topology. The key point is that, even in the presence of the magnetic
carpet, the structure of the coronal field is determined by the
balance between the Lorentz force and gas pressure. This means that, in
the low-beta corona where the field dominates, the time-averaged
field line geometry must resemble that of a fully closed potential or,
more generally, a force-free-equilibrium field. Note that this
constraint holds {\it even in the open field regions}.  Field lines
can deviate significantly from the potential/force-free state only at
heights where the gas pressure becomes of order the magnetic
pressure. Therefore, in those regions where the force-free field lines
never reach large heights, the field must be closed irrespective of
any ``noise'' due to the presence of the carpet. The occurrence of
open field lines in such a region would require a large deviation from
the geometry of the force-free state, which would violate basic force
balance.

The second assumption of the quasi-steady model is that there are no
long-lived current sheets in the closed field corona. (In this paper
we use the term ``current sheet'' to refer exclusively to a true
discontinuity in the field, not to merely a large concentration of
electric current.)  This second assumption is also likely to be valid,
because the corona is low beta and appears to be evolving slowly
(except during CMEs and flares), so the field must be in a fairly
robust equilibrium. It has been argued by several authors that current
sheets are not likely to exist in the force-free corona
\citep{vanbal85, ska87}. Note that the small-scale, transient currents
required by many models of coronal heating \citep[e.g.,][]{parker83}
are allowed, but are not expected to influence the global
corona. Furthermore, long-lived volumetric currents such as those
induced by photospheric shearing and twisting motions are also allowed
and can be arbitrarily large, as long as there are no true
discontinuities in the magnetic field. Of course in the source surface
models there are no currents whatsoever, the field is assumed to be
potential, but volumetric currents are almost certain to be present
in the MHD models.

\section{Coronal Hole Uniqueness}

Although the assumptions of smooth currents and a quasi-steady corona
seem innocuous, they actually place severe restrictions on coronal
hole topology. In particular, we claim that they imply the following
``uniqueness conjecture'':\\ {\it Every unipolar region on the
photosphere can contain at most one coronal hole.}

\noindent We do not have a rigorous mathematical proof of this
conjecture that holds for all possible topologies; instead, we present
below compelling arguments that it should hold for observed solar
conditions, as long as the assumptions above are valid. In particular,
it should hold for the quasi-steady, source-surface and MHD models. We
believe, however, that the conjecture is valid in general.

\subsection{Bipolar Topology}

To clarify our arguments for {\it uniqueness}, consider first the
simplest possible coronal topology: a bipolar field with one coronal
hole in each unipolar region. An analytic source-surface solution for
such a field can be obtained by using the method of images
\citep{jackson62}. For a dipole $\vec{d}$ located at a point
$\vec{r_d}$ inside the Sun, and a source surface at radius $R_S$, the
magnetic field potential, $\vec{B} = - \nabla \Phi$ is given by:
\begin{equation}
\Phi = \frac{\vec{d} \cdot (\vec{r} - \vec{r_d})}{| \vec{r} - \vec{r_d}|^3}
       - \frac{R_S r_d^3 \vec{d} \cdot (R_S^2 \vec{r} - r^2 \vec{r_d})}
              {| r_d^2 \vec{r} - R_S^2 \vec{r_d}|^3}.
\end{equation}
It is straightforward to verify that $\Phi = 0$ at the source surface,
$r = R_S$, so the field is purely radial there. Equation (1) is highly
useful, because one can build up a field of arbitrary complexity
simply by adding more dipoles. For the simplest possible case of a
single dipole at Sun center pointed along the vertical axis, Equation
(1) reduces to:
\begin{equation}
\Phi = \frac{(\vec{d}\cdot\vec{r})({R_S}^3 - r^3)}{{R_S}^3 r^3}
\end{equation}
and is shown in Figure 1a for $R_S = 2.5$ with the solar radius
normalized to unity. 

Topologically, the magnetic field can be considered to define a
mapping that connects the points along any field line in the
volume. This mapping is given by integrating the equations for a field
line:
\begin{equation}
\frac{dx}{ds} = \frac{B_x}{B}, \quad \frac{dy}{ds} = \frac{B_y}{B}, 
\quad {\rm and} \quad \frac{dz}{ds} = \frac{B_z}{B},
\end{equation}
where $s$ is distance along a field line and $x, y$, and $z$ are the
coordinates of the field line at the point $s$. It is apparent from
these formulas that, in order for the mapping to be discontinuous,
i.e., for field lines to``split", either one of the components $B_x,
B_y$, or $B_z$, must be discontinuous, which implies the presence of a
singularity in the current (a current sheet), or the magnitude $B$
must vanish. All the arguments below follow from this straightforward,
but very powerful result: {\it In the absence of singular
currents, magnetic field lines can split only at locations where the
field vanishes, such as true null points}. Although we will use this
result only in the context of solar coronal magnetic fields, we
emphasize that it holds in general. For example, the classical model
for the magnetosphere, a dipole embedded in a background field,
exhibits the topological feature that the field lines split only at
the magnetopause and distant-tail null points \citep{cowley73,
stern73, lau90}.

Note that the field of Fig. 1a does vanish at the null line located on
the equator of the source surface and, in fact, the field lines do
split there. (For a source surface solution the null is of the
X-type, whereas for a solution with a solar wind it would be of the
Y-type; but this difference is irrelevant to our argument.)  Equations
1 - 3 confirm our expectation that the mapping must be discontinuous
across a coronal-hole boundary. This boundary is a true separatrix
surface, just like the well-known fan surface described below. However
the mapping defined by Equation (2) is continuous everywhere else, in
particular, in the closed field region. Since there are no currents in
the closed field corona of Fig. 1a, the field line mapping is
especially simple; but, even if photospheric flows were applied that
greatly deformed the polarity distribution and that generated
strong current in the corona, the field line mapping would remain
continuous as long as the flows were smooth \citep{ska87}. Note that
we allow for the possibility of quasi-separatrix layers (QSL), where
the mapping exhibits large gradients \citep{titov02} as long as it is
not truly discontinuous.

The key point in the argument for {\it uniqueness} is that the field
lines in the closed-field region of Fig. 1a cannot split, because
neither of the necessary conditions is met: a simple bipolar region
contains no nulls and, by assumption, no current sheets are present
there. If the field lines cannot split, it is straightforward to
demonstrate that the positive and negative polarity regions of Fig. 1a
can each contain only one coronal hole. Assume that a second
disconnected hole exists, for example, in the northern hemisphere,
(Fig. 1b). This second hole must be fully inside the northern polarity
region, because as a direct consequence of the absence of current
sheets: \\ {\it A coronal hole boundary cannot intersect a polarity
inversion line (PIL)}.\\ This condition must be generally valid,
otherwise the heliospheric current sheet would extend all the way down
to the photosphere, violating our fundamental assumptions.  (Note that
to avoid confusion with magnetic null lines, we will use the
more-precise term PIL rather than the term ``photospheric neutral
line" that is also commonly used.) Since the coronal holes do not
intersect each other or the PIL, there must exist an annulus of flux
in the closed region that completely encircles one of the holes but
not the other, as sketched in Fig. 1b. This annulus flux must map to a
negative region across the equatorial PIL. Because the open flux of
the coronal holes constitutes a barrier that extends to infinity, it
is evident from Fig. 1b that the flux of any such annulus must go
around at least one of the coronal holes in order to connect across
the PIL; i.e., the field lines must split in the closed corona region.
But this is not possible in the absence of current sheets; hence, a
second disconnected coronal hole is forbidden.

This simple but compelling argument implies several important
points. If open field is intermixed with the closed field, as in
the interchange model, the topology is equivalent to many small
coronal holes embedded in the larger scale closed field. Therefore, by
reversing our argument, we conclude that the interchange model
requires the presence of many current sheets in the
corona. Furthermore, the current sheets must be inherently transient,
because unlike the quasi-steady heliospheric current sheet, gas
pressure cannot maintain current sheets in the low-beta corona.

\subsection{Open Field Corridors}

Another important point is that the coronal hole topology of Fig 1b
requires only minor modification to make it agree with the {\it
uniqueness} hypothesis: the addition of a very thin corridor of open
field connecting the two holes, as illustrated in Fig. 1c. The
geometry of the corridor can be arbitrary as long as it connects the
holes, in which case it becomes impossible to find an annulus of
closed field that passes between the two holes, and field-line
splitting is no longer an issue. If the holes are connected, however,
only one continuous coronal hole exists in the northern hemisphere,
and {\it uniqueness} still holds. Note, however, that the corridor may
be below observable resolution limits, so that the corona would appear
to contain two disconnected holes, reconciling observations with our
hypothesis. We contend that such narrow open-field corridors are
likely to be present in the real corona, providing a natural
explanation for the well-observed phenomenon of seemingly disconnected
coronal holes \citep{kahler02}.

It is easy to find open-field corridors in the source surface models
of observed solar fields.  Figure 2a shows the flux distribution for
Carrington rotation 1922, and 2b shows the regions of open and closed
field as calculated by the standard source surface model for this flux
distribution. Although the polarity regions and the PILs are much more
complex than the single dipole of Fig 1, both systems exhibit one
dominant polarity in the north and one in the south with a PIL
separating them. There are also numerous small opposite-polarity
regions with their PILs separating them from the main polarities, but
none of these appears to contain a coronal hole. It should be noted
that the actual flux distribution at the photosphere used to calculate
the source surface model of Fig. 2b contains many more opposite
polarity regions than can be seen in Fig. 2a, which for ease of
viewing, shows the flux slightly above the photosphere. These opposite
polarity regions are responsible for the numerous, small closed-field
circular regions near the boundary of the coronal hole in Fig. 2b. In
fact, there are undoubtedly many more such regions throughout the
polar coronal holes on the Sun than can be observed with present
instrumentation, so a true coronal hole map must actually resemble a
``swiss cheese'' pattern. We will consider the effect of such opposite
polarity regions later in this paper.

In the northern hemisphere near the center of Fig. 2b, the arrow
points to a coronal hole that appears to be well separated from the
main polar hole, but still in the same polarity region. If so, this
would clearly violate {\it uniqueness}. In order to investigate this
``disconnected'' hole in detail, we calculate an analytic
approximation to the field of Fig. 2a.  Note that for Fig. 2a the
solution was calculated numerically on a fixed grid using a
finite-difference scheme to solve Laplace's equation:
\begin{equation}
\nabla^2 \Phi = 0,
\end{equation}
with $\Phi = 0$ at the source surface.  A finite-difference solution
of Equation (4) is convenient for deriving initial conditions to a
time-dependent code, but for examining the detailed topological
properties of the source-surface model, a numerical solution is not as
effective as an analytic one. Therefore, we have taken the magnetic
flux distribution of Fig. 2a and calculated its spherical harmonic
expansion out to large order $l = 51$. Of course, a finite order
expansion does not return the identical flux distribution on the
boundary as in Fig. 2a, but this is not significant. Our only
requirement is that the flux distribution be approximated sufficiently
accurately that the ``disconnected'' hole is preserved.

Given the coefficients for the boundary-flux expansion, $C_{lm}$, we
can then write down the exact source-surface solution in the domain
\citep[e.g.,][]{wang92},
\begin{equation}
\Phi = \sum_{l=1}^{L} \sum_{m=-l}^{l} C_{lm}
\frac{Y_{lm}(\theta,\phi)} {r^{l+1}} \frac{(r^{2l+1} - R_s^{2l+1})}{(l
+ (l+1)R_s^{2l+1})},
\end{equation}
where, as in Equation (2), the solar radius is normalized to unity. The
advantage of this formulation is that, in principle, we can use the
analytic solution given by Equation (5) to determine the field line
mapping with arbitrary accuracy. (In practice, however, the
computational time required may be prohibitive.)  The other important
advantage of the expansion above is that simply by using different
values for the order of the expansion, $L$, we can investigate the
effect of applying different levels of smoothing to the photospheric
flux distribution.

We find that, for spherical harmonic solutions with order ranging up
to $L = 51$, all of the detailed open field structure evident in
Fig. 2b disappears except for the ``disconnected'' hole in the
north. Evidently, this hole is a robust feature of the photospheric
polarity distribution. Fig. 3 presents results from the $L = 31$
solution.  The first panel (Fig. 3a) shows the domain used for
plotting, along with the PIL on the photosphere (thin blue line) and
the PIL on the source surface (thick blue line). The photospheric
polarity is predominantly positive in the region shown, the north pole
is all positive, but near the center of the region, there is an
extended tongue of negative polarity oriented east-west. The
``disconnected'' hole of Fig. 2b lies south of this negative
tongue. The next panel (Fig. 3b) shows two sets of open field lines
(red and green) traced from the source surface down to the
photosphere. The starting footpoints for the two sets are two lines of
constant longitude separated by approximately $20^{\circ}$ in longitude
on the source surface. Each set begins with a field line whose
starting point is very close to the source surface PIL and whose end
point on the photosphere lies in the ``disconnected'' hole, south of
the negative polarity tongue. It is evident from Fig 3b that the field
maps a line of constant longitude on the source surface to a line on
the photosphere that passes from the small ``disconnected'' hole,
around the tongue and into the main northern polar hole.

If the small hole were truly disconnected, then the field-line mapping
would be discontinuous, and the line on the photosphere defined by
each field-line set would contain a break. Figs 3c and 3d show closer
views, (and from different perspectives), of the field line mapping at
the photosphere.  We note that the mapping is continuous, but as it
passes around the tongue-like PIL, the mapping develops very large
gradients there. We had to increase the density of starting footpoints
on the source surface by two orders of magnitude in order to find
field lines that map to locations near the tip of the tongue. For
expansions of significantly higher order, such as $L = 41$, the
negative polarity tongue becomes more extended, and the mapping
develops such large gradients that it becomes impossible with our
numerical integration routine to find open field lines near the tip of
the tongue by tracing downward from the source surface. To find such
field lines one needs to begin the trace near the photosphere. This
region of the field-line mapping can be considered to be an extreme
example of a QSL. The key point, however, is that the field-line
mapping is indeed continuous for all values of $L$, implying that a
narrow corridor of open field must connect the southern hole to the
main polar hole. Consequently, there is only one hole per unipolar
region, in agreement with {\it uniqueness}.

As calculated from the spherical harmonic solution with $L = 31$, the
corridor has extremely narrow photospheric width, of order only 10 km
at some locations.  Note that the photospheric footpoints of the two
field line sets are indistiguishable in Fig 3d, whereas the
source-surface footpoints are separated by scales of order the solar
radius. Such a small scale for the corridor width calls into question
the quasi-steady assumption. A corridor of this scale on the Sun would
be highly dynamic with field constantly opening and closing in
response to small changes in the photospheric flux. But as long as the
seemingly disconnected hole is present, then on average, a continuous
corridor must exist. Note that additional, wider corridors can clearly
be seen in the upper left of Fig. 2b. If enough corridors are present,
then on a large scale the open and closed field will appear to be
intermixed, so that near the coronal hole boundary the structure of
the quasi-steady models may begin to resemble that of the interchange
model. Furthermore, since they are likely to be continually dynamic
the corridors could become an important source of the slow wind. This
issue clearly needs further study.

\subsection{Multipolar Topology}

The argument above was developed for a simple bipolar magnetic
topology. The real corona, as seen in Fig. 2, usually contains other
large-scale structures, such as active regions. These add topological
complexity to the coronal field, especially null points where
field lines do split. Therefore, the next step in the proof for {\it
uniqueness} is to consider the effect on the arguments above of adding
an active-region bipole to the topology of Figure 1. If the active
region merely distorts the equatorial PIL, as in the well-studied case
of May 12, 1997 \citep{arge04}, the topology of the closed field is
still bipolar, and the annulus argument above holds. Similarly, if the
active region emerges inside one of the open field regions, then the
only effect is to produce a small closed-field region inside one of
the polar holes, which again has no effect on the topology of the main
closed region or on our argument. A significant change in topology
occurs only if the active region produces a new PIL in the closed
field photosphere, as in the field of Fig. 4a, where we have 
added a new low-latitude dipole source to Eq. (1). The potential
is now given by:
\begin{equation}
\Phi = \Phi_0 + \frac{\vec{d_1} \cdot \vec{r}}{| \vec{r} - \vec{r_1}|^3}
       - \frac{R_S^3 r_1^3 \vec{d_1} \cdot \vec{r}}
              {| r_1^2 \vec{r} - R_S^2 \vec{r_1}|^3},
\end{equation}
where $\Phi_0$ is the potential of Equation (2), $\vec{d_1} = (0,
A, 0)$, and $\vec{r_1} = (.9, 60^\circ, 0)$. In other words, a dipole
pointing due north is placed at a latitude of $30^\circ$ and a depth of
$0.1 R_\odot$.

The addition of the active-region dipole produces a new PIL separating
the negative active-region spot from its positive surroundings.  We
will use the term {\it nested} polarity region to refer to a
configuration like the negative spot, which is wholly surrounded by a
larger opposite-polarity region. The fact that some of the
surroundings are in the form of a strong positive spot just south of
the strong negative spot is not important to the topology. Associated
with the active-region PIL is a null point in the corona, along with
the usual dome-shaped fan separatrix surface and pair of spine lines
\citep[e.g.,][]{greene88, lau90, ska90, priest96}. The intersection of
the fan with the photosphere forms a closed separatrix curve defining
the boundary between the positive flux connecting across the
active-region PIL and that connecting across the equatorial PIL. Note
that the field lines of the fan and spines all connect to the null and
split there; consequently, the mapping is discontinuous at the fan and
spines. The magnetic field of Fig. 4a is simply the well-known
embedded bipole, the most likely topology if there are two PILs (three
polarity regions) on the photosphere \citep{ska98}.

The only other possible topology for a two-PIL photosphere is that of
a ``bald patch'' in which the null point occurs below the surface and
the magnetic field over part of the nested PIL is concave up
\citep[e.g.,][]{titov93}. Bald patch topologies can occur only if the
nested polarity region is small; therefore, they are unlikely to play
a significant role in determining the large-scale solar field, such as
the coronal hole topology. On the other hand, they have interesting
implications for coronal dynamics, because we do not expect bald patch
topologies to survive in open field regions (Antiochos \& Mueller 2007
{\it in preparation}). Determining the evolution of bald patch
topologies is problematic, however, because simple line-tied boundary
conditions cannot be used wherever the coronal field is concave up at
the photosphere \citep{ska90, karpen90}. Consequently, we will not
consider bald patch topologies further in this paper.

It should be emphasized that the null-point topology of Fig. 4a is
observed to be a generic feature of coronal magnetic fields. It was
immediately seen by Skylab, where it was referred to as a ``fountain''
region \citep{tousey73, sheeley75}, and by Yohkoh where it was
referred to as an ``anemone'' region \citep{shibata94,
vourlidas96}. As demonstrated by numerous extrapolations of observed
photospheric fields, it is ubiquitous throughout the Sun on a broad
range of scales \citep[e.g.,][]{aulanier00, fletcher01, luhmann03,
urra07}. Of course, the true solar field is almost always more complex
than that of a single active-region bipole. However, if the
photospheric flux consisted of clearly separated embedded bipoles,
each with its own PIL, then the topology would simply be that of a
collection of non-intersecting fans and spines. Even for complex
active regions, we expect that such regions also appear mainly
bipolar, on the large scale that is important for determining the
global field. Therefore, if {\it uniqueness} holds for the topology
of Fig. 4a, it is likely to hold in general for all observed solar
fields.

We now add a disconnected coronal hole to the system shown in Fig. 4a,
and consider the implications of the null-point topology for our
uniqueness conjecture. If the separatrix curve of the nested polarity
does not intersect a coronal hole boundary as in Fig. 4b, then it is
always possible to find an annulus of closed flux surrounding either
hole that maps across the equatorial PIL. In this case the presence of
the nested polarity is irrelevant, and the field of the annulus must
split around one of the holes, which is disallowed. Thus, our earlier
argument for uniqueness applies without change. But what happens if
the active region moves or expands, such that its associated
separatrix curve intersects the coronal holes as sketched in Fig. 4c?
In this case there is no annulus of flux closing across the equator
that encircles only one of the holes -- all such annuli encircle both
holes, so that no splitting of the field is required. Furthermore, any
annulus of closed flux that encircles only one of the holes must cross
the nested-polarity separatrix curve, in which case the annulus flux is
allowed to split at the null point.

Although a rigorous treatment of the topology implied by Fig. 4c
requires a fully dynamic calculation, we can gain useful insight by
considering the topology predicted by a sequence of static models in
which a nested-polarity separatrix curve approaches a coronal hole
boundary. Figure 5 shows the source-surface model topology for two
slightly different positions of the embedded dipole.  The dipole in
Fig. 5a is located only $1^{\circ}$ south of that in 5b. The changes in
location and shape of the PIL and the separatrix curve between
Figs. 5a and 5b are imperceptible, but the open-field topologies are
dramatically different. The coronal hole boundary passes completely
north of the fan separatrix curve in 5a and completely south in 5b, so
the separatrix and the coronal hole never actually intersect. A point
of clarification is that, when the nested polarity is inside the
coronal hole as in 5b, the fan separatrix, itself, can be considered
to define a coronal hole boundary, because it separates the flux that
closes across the nested polarity PIL from the surrounding open
flux. Note, however, that this type of coronal hole boundary is
confined to low heights, well below the source surface, because the
fan field lines all connect to an X-type null low in the corona.  For
the issue of {\it uniqueness}, the only coronal hole boundaries that
are important, and that we will refer to, are those that connect to
the source surface.

Figure 5 shows the change in topology for a one degree shift in the
embedded dipole's position, but we find the same result no matter how
small the shift: either the separatrix curve is fully inside the
closed field region as in 5a, or it is fully inside the coronal hole
as in 5b. The implication of the quasi-steady model, therefore, is
that the open field topology undergoes a discontinuous jump as a
nested polarity region approaches a coronal hole boundary. This result
may seem unphysical, but it follows inevitably from the fan - spine
topology of Fig. 4a, in which the spine lines split at the null to
form the fan. The actual amount of flux in the spines and fan is a set
of measure zero, so this picture is to be taken in the sense of a
limit. The relevant point, however, is that any arbitrarily small but
finite flux bundle enclosing the outer spine maps to an arbitrarily
narrow but finite-width annulus on the photosphere surrounding the
separatrix curve. Therefore, if the outer spine is closed (connects to
the photosphere), then the fan is surrounded by closed flux and the
nested polarity must be considered to be in the closed field
region. For this case the null is inside the closed field region as in
Fig. 5a. Conversely, if the outer spine is open (connects to the
heliosphere), the fan must be surrounded by open flux and the nested
polarity along with the null is in the coronal hole, as in
Fig. 5b. The case where the outer spine is exactly on the coronal hole
boundary corresponds to the fan being surrounded by an open region of
vanishing width -- a singular case that can be neglected.

We conclude that the configuration shown by Fig. 4c is impossible,
because:\\ {\it A nested polarity region must be surrounded by either
all open or all closed field.}\\ In other words, a separatrix curve
cannot intersect a coronal hole boundary and, consequently, the
correct topology for the system of Fig. 4c must actually be that shown
in Fig. 5c: a seemingly disconnected coronal hole connected by an
open-field corridor. But if separatrix curves and coronal hole
boundaries cannot intersect, then the annulus argument can always be
applied and {\it uniqueness} holds even in a multipolar topology like
that of Fig. 4.

Since Fig. 5 shows only a sequence of potential-field states, a
critical question that immediately arises is whether a true dynamical
evolution will be compatible with this sequence.  When a bipole is
convected by photospheric flows toward a coronal hole, we expect that
the null point will deform into a current sheet similar to the classic
\citet{syrovatskii81} theory, and that magnetic reconnection will
occur between the spine and external flux. Such reconnection has been
observed in many numerical experiments
\citep[e.g.,][]{parnell04,pontin07}. This spine reconnection will act
so as to exchange the outer spine with external flux, effectively
moving the spine flux toward the coronal hole boundary (and also
destroying the current sheet). When the spine reaches the coronal hole
boundary, reconnection between open and closed flux will move the
outer spine into the open field region. Once inside the coronal hole,
any subsequent motion will result in further interchange reconnection,
as has been proposed in models for heating the solar wind and coronal
hole plumes \citep{parker92, axford92, deforest98}. Although, this
scenario awaits verification with fully time-dependent calculations,
which are still in progress \citep{ska06}, it seems clear, however,
that magnetic reconnection can readily produce the evolution implied
by the source-surface solutions of Fig 5.

\section{Coronal Hole Nesting}

An important feature of the topology of Fig. 5b is the very thin
(below the resolution of the Figure) open-field corridor, attached to
the polar coronal hole at both ends, that passes around the south side
of the fan, and right over the center of the strong positive spot
there. Although the presence of the corridor implies that the coronal
hole is no longer simply connected, it is still a single, unique
coronal hole. Such corridors should form naturally on the Sun any time
a bipolar region moves into a coronal hole. In fact, as noted in
Section 3.2, such corridors can be seen at the edges of the polar
coronal holes in Fig. 2b.

Open field corridors play a critical role in reconciling multipolar
topologies with our {\it uniqueness} conjecture. Let us return to the
topology of Fig. 4a, in which the active-region separatrix curve is
well removed from the polar coronal-hole boundary, and consider the
effect of opening a coronal hole inside the nested polarity, as
sketched in Figs. 6a and 6b. This should be allowed by the {\it
uniqueness} conjecture, because there would still be only one coronal
hole per unipolar region on the Sun. But the annulus argument forbids
the topology of Figs. 6a and 6b on two counts. First, consider any
annulus of closed flux that surrounds the inner spine. The inner spine
maps to the whole fan surface, which surrounds the nested polarity,
therefore, any annulus surrounding the spine must map to an annulus
surrounding the nested polarity.  But in order to pass around the
embedded polarity coronal hole, any such annulus of closed flux would
have to split, which is not allowed. This problem is easily taken care
by insisting that:\\ {\it Any coronal hole that opens inside a nested
polarity must encompass the spine.}\\ This conclusion has important
implications for models of CME initiation such as the breakout model
\citep{ska99}.  It predicts that prior to eruption the inner spine
should appear to move toward that part of the sheared PIL that
eventually erupts. Furthermore, the amount of energy available for
eruption will depend on how much the spine can move
\citep{ska99,devore05}.

The second application of the annulus argument leads to a more
surprising conclusion. Any annulus of closed flux that surrounds the
polar coronal hole of Fig. 6a would clearly have to split even if the
embedded coronal hole had a permissible, open-spine topology. It would
appear, therefore, that the annulus argument implies not just one hole
per unipolar region, but only one hole per polarity (i.e., only two on
the whole Sun), which cannot be right. The resolution to this
conundrum is illustrated by Fig. 6c -- the formation of an open-field
corridor that passes completely around the nested polarity region. In
Fig. 6c, field line splitting is no longer a problem, because any
annulus of closed flux that surrounds the main coronal hole must also
surround the active-region hole. Also, any annulus of closed flux
inside the nested region that surrounds the nested hole closes
completely within the closed flux region of the nested polarity.
These arguments imply that multiple coronal holes are, indeed,
possible, but they must obey the following {\it nesting} conjecture:\\
{\it Coronal holes of nested polarity regions must themselves be
nested.}

Although, the {\it nesting} conjecture imposes a powerful constraint
on coronal hole topology, its importance for observed solar fields is
uncertain. Nested coronal holes imply the presence of small conical
current sheets in addition to the main heliospheric current sheet. In
situ measurements usually indicate a single current sheet in the
heliosphere, implying that there are only two coronal holes on the
Sun. If so, then the issue of nested coronal holes becomes moot. Note
that the nested polarity flux must be large in order to obtain a
nested coronal hole within the context of the source surface model, in
other words, a large active region far from the equator. This
combination is rarely observed on the Sun, but still, it would be
intriguing to search for any such nested holes in the published source
surface maps, and then to search for their current sheets in the
heliosphere.

Furthermore, the source surface model is likely to underestimate the
occurrence of nested coronal holes, because this model does not
include force balance between field and plasma. The topology of
Fig. 5b contains a simple X-type null on the fan surface separating
open and closed field. If plasma is added to the configuration of
Fig. 5b, a difference in gas pressure will develop between the
confined plasma inside the fan and the unconfined, solar wind plasma
outside, just as there is a well-observed difference in the gas
pressure between solar closed-field regions and coronal holes. For low
beta, a gas pressure gradient across the fan or any other magnetic
surface can readily be balanced by a small magnetic pressure gradient
there; but, this is not possible near the null where the beta becomes
infinite. It seems, therefore, that the plasma pressure would strongly
deform the field near the null, and in some cases, may force open a
finite region of flux around the inner spine even though the source
surface model predicts no nested coronal hole. Of course, the nested
polarity regions on the Sun are likely to be evolving via flux
emergence or cancellation and photospheric motions. In fact, several
models for accelerating the wind \citep{parker92, axford92} and for
forming polar plumes \citep{deforest98} invoke this process of
interchange reconnection between the closed flux of an embedded bipole
and surrounding open field.  Therefore, it may be that any small
nested holes are masked by the reconnection and dynamics.  There are
bound to be cases, however, where large quasi-static bipoles appear
inside coronal holes. We predict that, at least for these cases, there
would be a significant and possibly observable difference between the
topology predicted by the source surface and the MHD model.

One situation in which the {\it nesting} conjecture is quite likely to
play an important role is in long-lived dimming regions formed by
CMEs. The dimming regions are believed to be transient coronal holes
where the magnetic field has been forced open by a CME
\citep{thompson00}. Since they are transient, it is not clear that our
arguments above apply. If the holes are sufficiently long-lived (time
scales of tens of hours), however, the quasi-steady assumption may
still be valid. If so, then we can make two predictions on such
``not-too-transient'' holes. First, any dimming region that forms
inside an active region PIL must encompass the inner spine. This
prediction probably is difficult to test, because the inner spine is
not easily observed \citep{aulanier00}. Second, the formation of such
a dimming region must be accompanied by the formation of a transient
coronal hole (possibly a very narrow open-field corridor) surrounding
the PIL. This latter prediction may well be testable in some
well-observed CMEs.

\section{Discussion}

Let us summarize our main findings and predictions on coronal hole
topology. First, we list two supporting ``lemmas'':\\ {\it A coronal
hole boundary cannot intersect a polarity inversion line.}\\ This
statement is rigorously valid for the quasi-steady models, because the
heliospheric current sheet cannot extend down to the photosphere.  In
the interchange models, however, open flux is presumably free to
diffuse across PILs, which emphasizes the striking difference between
the two models.

\noindent {\it A nested polarity region must be surrounded by
either all open or all closed field.}\\ The prediction from this
result is that coronal hole boundaries undergo discontinuous jumps in
response to bipolar regions entering or exiting the holes.

Next, application of the annulus-of-closed-flux argument leads to our
two main results, the {\it uniqueness} and {\it nesting}
conjectures:\\ {\it Every unipolar region on the photosphere can
contain at most one coronal hole.}\\ We predict that seemingly
disconnected holes are actually connected by observationally
unresolved open field corridors. These corridors are likely to be
dynamic, with the field continuously opening and closing in response to
photospheric motions and flux emergence or submergence. Furthermore, we
expect such corridors to be ubiquitous at the boundaries of coronal
holes, causing these boundaries to have a fractal-like and inherently
dynamic structure (see Fig. 2b). Consequently, the corridors may be an
important source of the slow wind.

\noindent{\it Coronal holes of nested polarity regions must themselves
be nested.}\\ The prediction is that if a coronal hole develops in an
active region (i.e., a nested polarity region), then the polar coronal
hole will grow to surround the nested polarity. This may hold even for
transient coronal holes associated with CMEs. Related to this
conjecture is the corollary: \\{\it Any coronal hole that opens inside
a nested polarity must encompass the spine.}

We emphasize that the statements are still conjectures, even for the
quasi-steady models. The arguments presented in this paper used only
basic topologies for the coronal field. It may well be that one can
find counter-examples, especially in systems with special symmetries
so that structures such as null lines or null surfaces appear in the
corona. On the other hand, the coronal magnetic field is generally
observed to have smooth structure without evidence for such
topological pathologies. Note also that topologies such as null lines
are structurally unstable, in general, so they would exist only as
transient structures. Therefore, if our conjectures are valid for the
topologies discussed above, it seems likely that they will hold for
most observed solar magnetic fields.

We further emphasize that, for application to the Sun, all our results
depend on the underlying assumption that the large-scale corona can be
considered to be in a quasi-steady equilibrium state, as in the source
surface and MHD models. If time-dependent effects dominate instead,
as in the interchange models, then the statements above are unlikely
to be valid. Consequently, observational testing of our conjectures
may be the most effective method for determining the correct theory
for the solar-heliospheric magnetic field.

\acknowledgments

This work has been supported by NASA, ONR and the NSF SHINE
Program. The work has benefited greatly from the authors'
participation in the NASA TR\&T focused science team on the
solar-heliospheric magnetic field. Discussions with the team leader,
Thomas Zurbuchen, were especially helpful. The work of Z. Mikic was
supported, in part, by the NASA TR\&T focused science team on the
CME-ICME connection.




\clearpage

\begin{figure}
%%\epsscale{.95}
\plotone{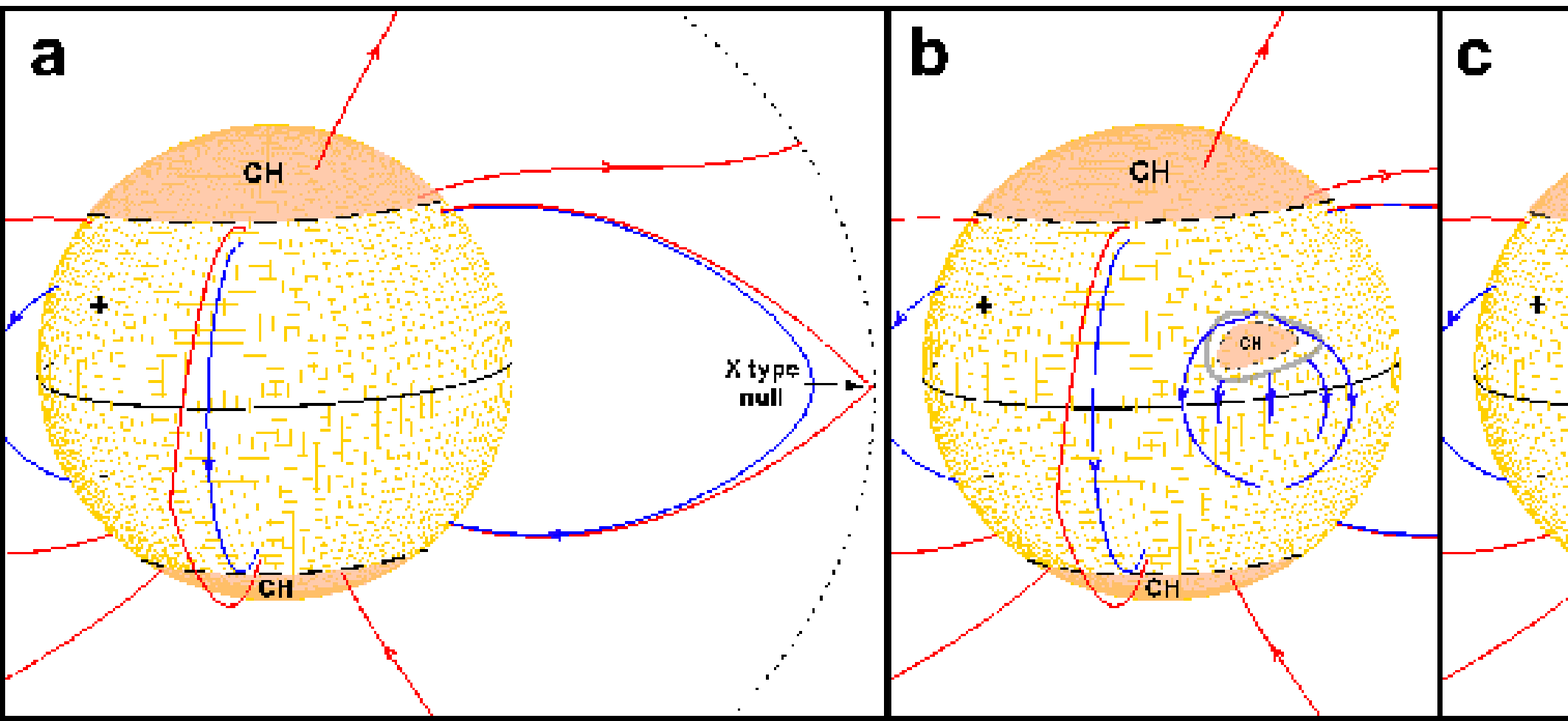}
\caption{ (a) Source-surface solution for a bipolar coronal field with
two coronal holes indicated by the darker areas. The solid black line
denotes the PIL at the equator, while the dashed black lines indicate
the boundaries of the coronal holes. Red and blue solid lines denote
open and closed field lines, respectively. (b) A hypothetical
disconnected coronal hole is drawn in the northern hemisphere of
(a). An example of an annulus of closed flux surrounding one of the
holes is shown in grey, with hypothetical field lines that close
across the equator in blue. (c) An example of an open field corridor
connecting the coronal holes of panel (b).
\label{f1}}
\end{figure}

\clearpage
\begin{figure}
%%\epsscale{.95}
\plotone{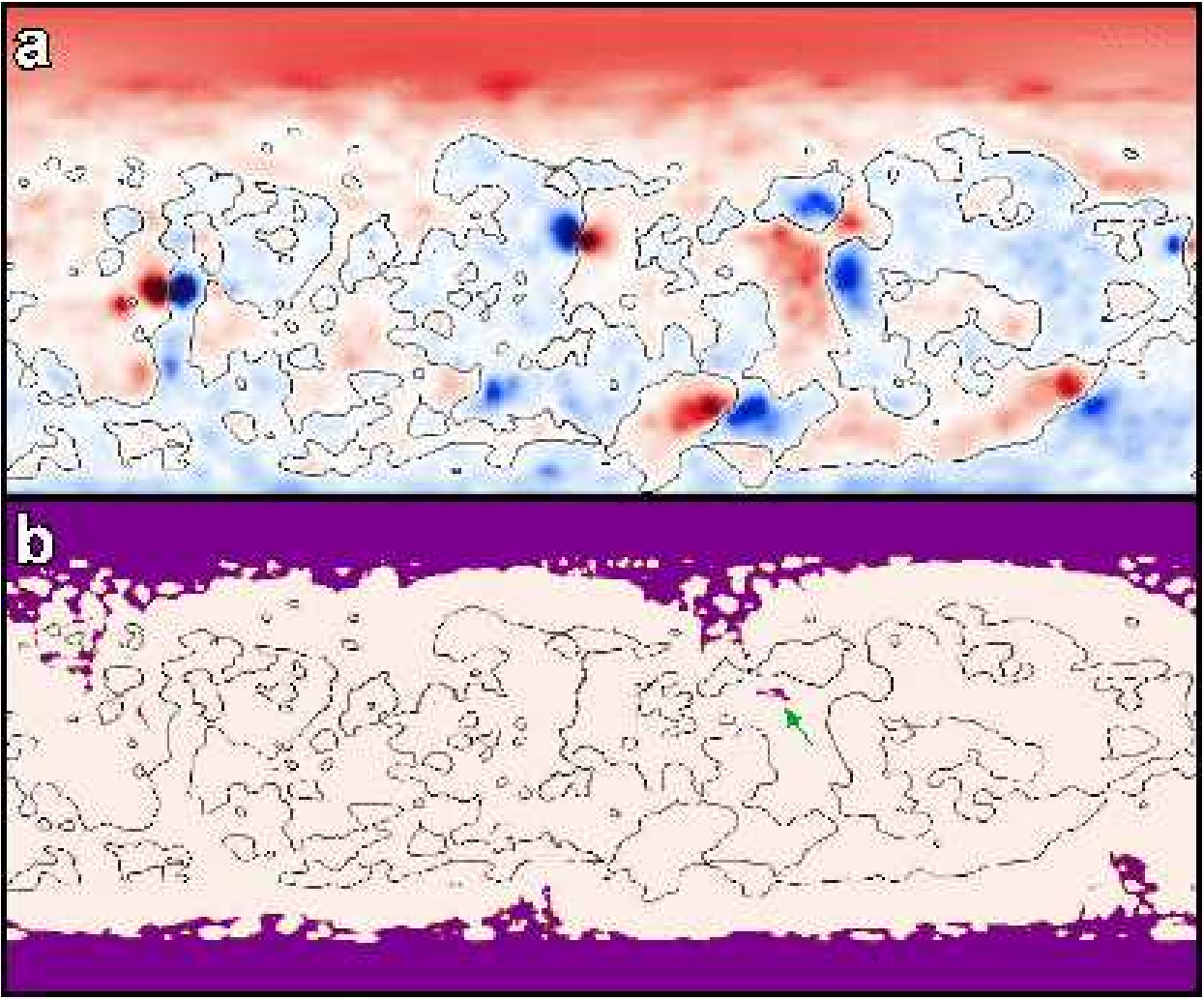}
\caption{ (a) Photospheric polarity from MDI magnetograms for
Carrington rotation 1922. (b) Source surface solution showing open
(purple) and closed (off white) regions along with the photospheric
PIL (black lines). The apparently disconnected hole is marked by an arrow.
\label{f2}}
\end{figure}

\clearpage

\begin{figure}
%%\epsscale{.95}
\plotone{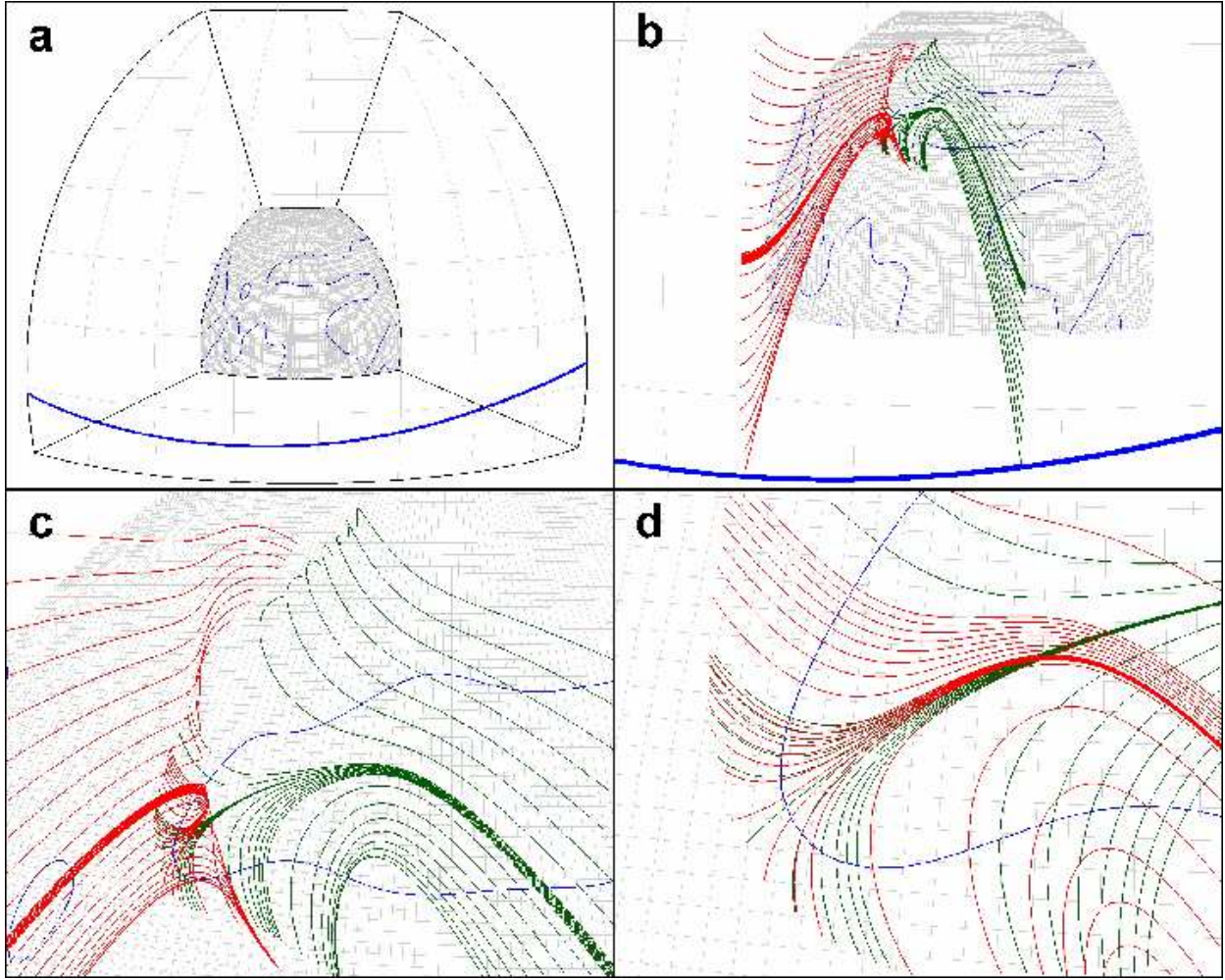}
\caption{ (a) Plotting domain for a spherical harmonic expansion
solution with $L = 31$. The blue contours correspond to the PILs on
the photosphere and on the source surface. (b) Sets of field lines
(red and green lines) traced from lines of constant longitude on the
source surface down to the photosphere. (c) View of the photospheric
footpoint positions of the field lines of the previous panel. (d) A
close-up of the region near the tip of the negative polarity tongue
showing that the footpoints of the red and green field lines become
unresolvably close in this region.
\label{f3}}
\end{figure}

\clearpage

\begin{figure}
%%\epsscale{.95}
\plotone{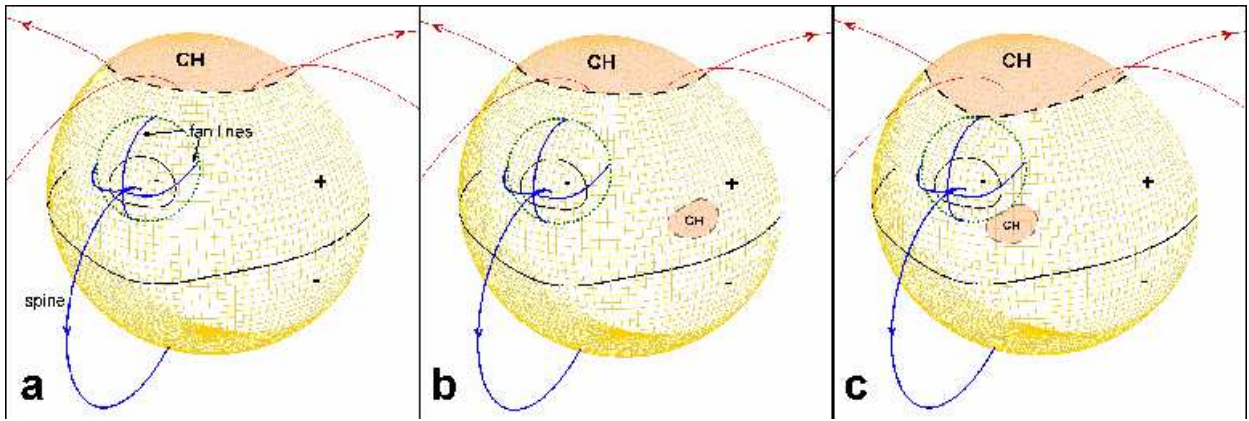}
\caption{(a) Source-surface solution for a bipolar active region
embedded in the global dipole of Fig. 1. For clarity, the southern
coronal hole is not shown. A 3D null occurs at the point of
intersection of all the blue field lines. The blue line closing into
the negative south is the outer spine; the line closing into the
negative spot is the inner spine. The other 4 lines lie on the fan
surface. The green dotted line indicates the separatrix curve, the
intersection of the photosphere and the fan surface.  (b) A
disconnected coronal hole in the topology of panel (a) that is
forbidden by the annulus argument. (c) A hypothetical topology in
which the nested-polarity separatrix curve intersects the two coronal
holes, thereby circumventing the annulus argument and contradicting
{\it uniqueness}.
\label{f4}}
\end{figure}

\clearpage

\begin{figure}
%%\epsscale{.95}
\plotone{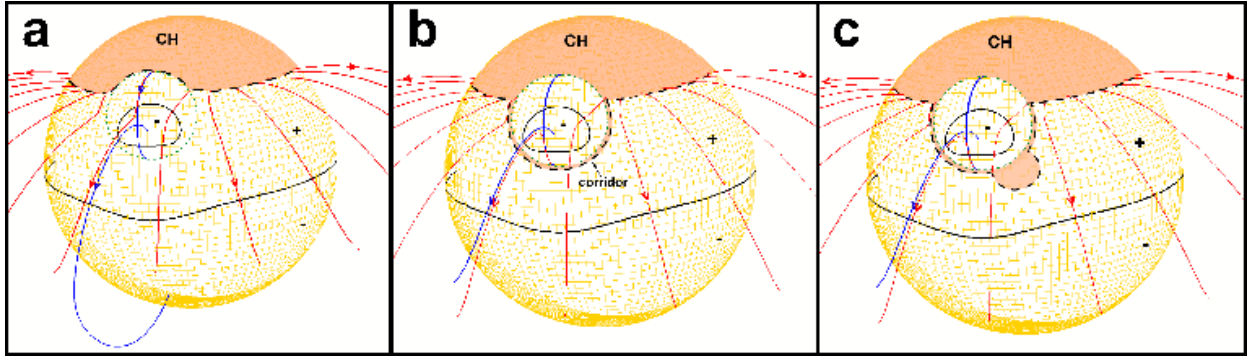}
\caption{(a) Source-surface solution as in Fig. 4a, but for an active
region dipole that is further north. (b) The source-surface solution
for the dipole moved just $1^{\circ}$ north from (a). The outer spine
is now open and the coronal hole boundary completely surrounds the
nested polarity region. An open-field corridor separates the closed
nested-polarity flux from the closed global flux. (c) The correct
topology for the system of Fig. 4c.
\label{f5}}
\end{figure}

\begin{figure}
%%\epsscale{.95}
\plotone{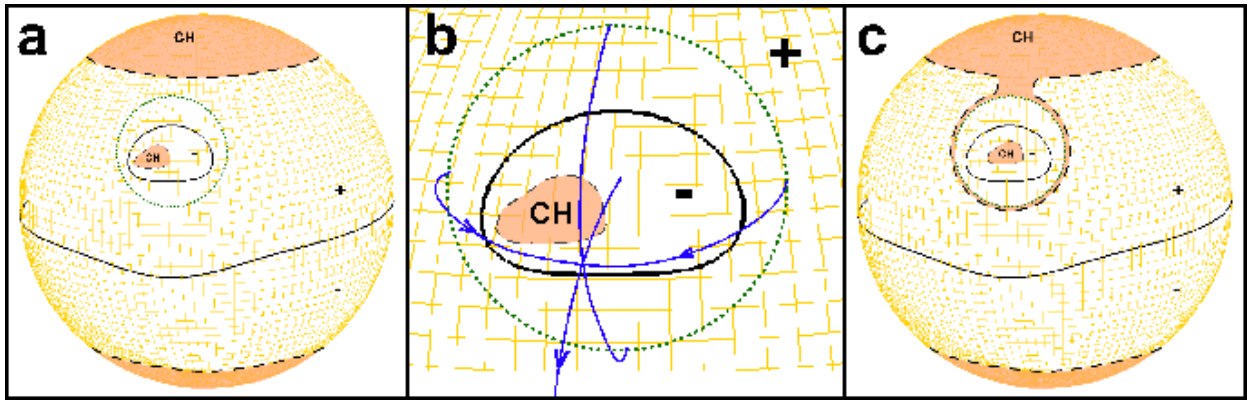}
\caption{ (a) Hypothetical topology satisfying {\it uniqueness} but
still forbidden by the flux annulus argument. (b) Close-up of this
topology inside the nested-polarity region. (c) Nested coronal hole
topology that is not forbidden.
\label{f6}}
\end{figure}

%% The following command ends your manuscript. LaTeX will ignore any text

\end{document}